\documentstyle[tighten,floats,12pt,prd,aps,eqsecnum]{revtex}

\input epsf
\begin{document}
\date{\today}
\title{Effective potential in Brane-World scenarios}

\author{Oriol Pujol{\`a}s
\\ \hspace*{1cm}}

\address{~IFAE, 
Departament de F{\'\i}sica, Universitat Aut{\`o}noma de Barcelona,\\
08193 Bellaterra $($Barcelona$)$, Spain}

%\address{$^{2}$~Yukawa Institute for Theoretical Physics, 
%Kyoto University, Kyoto 606-8502, Japan

\maketitle

\thispagestyle{empty}
\vspace{1cm}
\centerline{\bf Abstract}
\vspace{2mm}
\begin{abstract}
    We review the stabilization of the radion in the Randall--Sundrum model
    through the Casimir energy due to a bulk conformally coupled field.  We also
    show some exact self--consistent solutions taking into
    account the backreaction that this energy induces on the geometry.
\end{abstract}

\hspace*{1cm}
$~~~~~~~~~~~~~~~~~~~~~~~~~~~~~~~~~~~~~~~~~~~~~~~~~~~~~~~~~~~~~~~~~~~~~~~~~~~~$
UAB-FT-504 
\vspace{1cm}
%\newpage

\section{Introduction}

Recently, it has been suggested that
theories with extra dimensions may provide a solution to the hierarchy
problem \cite{gia,RS1}. The idea is to introduce a $d$-dimensional 
internal space of large physical volume 
${\cal V}$, so that the the effective lower dimensional Planck mass 
$m_{pl}\sim {\cal V}^{1/2} M^{(d+2)/2}$ is much larger than $M \sim TeV$-
the true fundamental scale of the theory. In the original
scenarios, only gravity was allowed to propagate in the higher
dimensional bulk, whereas all other matter fields were confined to live 
on a lower dimensional brane. 
Randall and Sundrum \cite{RS1} (RS) introduced a particularly attractive
model where the gravitational field created by the branes is taken into 
account. Their background solution consists of two parallel flat branes, one 
with positive tension and another one with negative tension 
embedded in a a five-dimensional Anti-de Sitter (AdS) bulk. In this model,
the hierarchy problem is solved if the distance between branes 
is about $37$ times the AdS radius and 
we live on the negative tension brane. More recently, 
scenarios where additional fields propagate in the bulk have 
been considered \cite{alex1,alex2,alex3,bagger}.

In principle, the distance between 
branes is a massless degree of freedom, the radion field $\phi$.
However, in order to make the theory  compatible with observations 
this radion must be stabilized \cite{gw1,gw2,gt,cgr,tm}. Clearly, 
all fields which propagate in the bulk will give Casimir-type contributions
to the vacuum energy, and it seems natural to investigate whether these
could provide the stabilizing force which is needed.
Here, we shall calculate the radion one loop effective potential 
$V_{\hbox{\footnotesize\it \hspace{-6pt} eff\,}}(\phi)$ due to conformally
coupled bulk scalar fields, although the result shares many features with other
massless bulk fields, such as the graviton, which is  addressed in
\cite{gpt}. As we shall see, this
effective potential has a rather non-trivial behaviour, 
which generically develops a local 
extremum. Depending on the detailed matter content, the 
extremum could be a maximum or a minimum, where the radion could sit. 
For the purposes  of illustration, here we shall concentrate 
on the background geometry discussed by Randall and Sundrum, although 
our methods are also applicable to other geometries, such as the one 
introduced by  Ovrut {\em et al.} in the 
context of eleven dimensional supergravity with one large extra dimension
\cite{ovrut}. This report is based on a work done in 
collaboration with Jaume Garriga and Takahiro Tanaka \cite{gpt}.

Related calculations of the Casimir interaction amongst branes have been 
presented in an interesting paper by Fabinger and Ho\v rava \cite{FH}. In 
the concluding section we shall comment on the differences between their 
results and ours.

\section{The Randall-Sundrum model and the radion field}

To be definite, we shall focus attention on the 
brane-world model introduced by Randall and Sundrum \cite{RS1}.
In this model the metric in the bulk is anti-de Sitter space
(AdS), whose (Euclidean) line element is given by 
\begin{equation}
 ds^2=a^2(z)\eta_{ab}dx^{a}dx^{b}=
     a^2(z)\left[dz^2 +d{\bf x}^2\right] 
   =dy^2+a^2(z)d{\bf x}^2.
\label{rsmetric}
\end{equation}
Here $a(z)=\ell/z$, where $\ell$ is the AdS radius.
The branes are 
placed at arbitrary locations which we shall denote by $z_+$ and
$z_-$, where the positive and negative signs refer to the positive and
negative tension branes respectively ($z_+ < z_-$).
The ``canonically normalized'' radion modulus
$\phi$ - whose kinetic term contribution to the
dimensionally reduced action on the positive tension brane is given by 
\begin{equation}
 {1\over 2}\int d^4 x \sqrt{g_+}\, g^{\mu\nu}_+\partial_{\mu}\phi 
    \,\partial_{\nu}\phi, 
\label{kin}
\end{equation}
is related to the proper 
distance $d= \Delta y$ between both branes in the following way \cite{gw1}
$$
\phi=(3M^3\ell/4\pi)^{1/2} e^{- d/\ell}.
$$
Here, $M \sim TeV$ is 
the fundamental five-dimensional Planck mass. It is usually assumed
that $\ell \sim M^{-1}$ . Let us introduce the dimensionless radion
$$
\lambda \equiv \left({4\pi \over 3M^3\ell}\right)^{1/2} {\phi} = 
{z_+ \over z_-}   = e^{-d/\ell},
$$
which will also be refered to as {\em the hierarchy}. 
The effective four-dimensional Planck mass $m_{pl}$ 
from the point of view of the negative tension brane is 
given by $m_{pl}^2 = M^3 \ell 
(\lambda^{-2} - 1)$. With $d\sim 37 \ell$, 
$\lambda$ is the small number responsible for the 
discrepancy between $m_{pl}$ and $M$.

At the classical level, the radion is massless. However, as we shall see, 
bulk fields give rise to a Casimir energy which depends on the interbrane
separation. This induces an effective potential $V_{\hbox{\footnotesize\it \hspace{-6pt} eff\,}}(\phi)$ which by
convention we take to be the energy density per unit physical volume on the
positive tension brane, as a function of $\phi$.
This potential must be added 
to the kinetic term (\ref{kin}) in order to obtain the effective action for 
the radion:
\begin{equation}
 S_{\hbox{\footnotesize\it \hspace{-6pt} eff\,}}[\phi]
 =\int d^4x\, a_+^4 \left[{1\over 2}g_+^{\mu\nu}\partial_{\mu}\phi\, 
         \partial_{\nu}\phi +
          V_{\hbox{\footnotesize\it \hspace{-6pt} eff\,}}(\lambda(\phi))
        \right].
\label{effect}
\end{equation}
In the following Section,  we calculate the contributions to 
$V_{\hbox{\footnotesize\it \hspace{-6pt} eff\,}}$ from
conformally invariant bulk fields.

\section{Massless scalar bulk fields}

The effective potential induced by scalar fields with arbitrary coupling to the
curvature or bulk mass and boundary mass can be addressed. It
reduces to a similar calculation  to minimal the coupling
massless field case, which is  sovled in \cite{gpt}, and correponds to bulk
gravitons. However, for the sake of simplicity, we shall only consider below the
contribution to $V_{\hbox{\footnotesize\it \hspace{-6pt} eff\,}}(\phi)$ from
conformally coupled massless bulk fields. Technically, this is much simpler
than finding the contribution from bulk gravitons and the problem of
backreaction of the Casimir energy onto the background can be taken into
consideration in this case. Here we are considering generalizations of the
original RS proposal \cite{alex1,alex2,alex3} which allow several fields other
than the graviton only (contributing as a minimally coupled scalar field).

A conformally coupled scalar $\chi$
 obeys the equation of motion
\begin{equation}
-\Box_g \chi + {D-2 \over 4 (D-1)}\ R\ \chi =0, 
\label{confin}
\end{equation}
% where $\Box_g$ is the d'Alembertian operator in the metric
% (\ref{rsmetric}) and $R$ is the Ricci scalar. 
% Here, we consider the case 
% of arbitrary odd spacetime dimension $D$, with branes of co-dimension one.
% By a simple reescaling of the the variable 
% $\chi \to \hat\chi = a^{(D-2)/2} \chi$, the equation of motion for $\hat\chi$
% becomes
\begin{equation}
\Box^{(0)} \hat\chi =0.
\label{fse}
\end{equation}
Here $\Box^{(0)}$ is the {\em flat space} d'Alembertian. It is customary to
impose $Z_2$ symmetry on the bulk fields, with some parity given. If we choose
even parity for $\hat\chi$, this results
in Neumann boundary conditions
$$
\partial_{z}\hat\chi = 0,
$$
at $z_+$ and $z_-$. 
The eigenvalues of the d'Alembertian subject to these conditions are
given by 
\begin{equation}
\label{flateigenvalues}
\lambda^2_{n,k}=\left({n \pi \over L}\right)^2+k^2,
\end{equation}
where $n$ is a positive integer, $L=z_{-}-z_+$ is the coordinate distance
between both branes and $k$ is the coordinate momentum parallel to the branes.
\footnote{If we considered an odd parity field, then we would impose Dirichlet
  boundary conditions, $\hat\chi(z_-)=\hat\chi(z_+)=0$, and the set of
  eigenvalues would be the same except for the zero mode, which only the even
  field has.}

Similarly, we could consider the case of massless fermions in the RS 
background. The Dirac equation,\footnote{
Here, $e^a_{\ n}$ is the f{\"u}nfbein, $n,m,\dots$ are flat indices, $a,b,\dots$ are
``world'' indices, and
$\gamma^n$ are the Dirac matrices.
The covariant derivative can be expressed in terms of the spin connection
$\omega_{an m}$ as
$\nabla_a=\partial_a+{1\over 2} \omega_{anm} \Sigma^{nm}$, where
$\Sigma^{nm}={1\over 4}[\gamma^n,\gamma^m]$ 
are the generators of the Lorentz transformations in spin $1/2$ representation.}
$$
\gamma^{n}e^a_{\ n}\nabla_a\,\psi=0.
$$ 
is conformally invariant
\cite{bida}, and the conformally rescaled components of the 
fermion obey the flat space equation (\ref{fse}) with Neumann boundary 
conditions. Thus, the spectrum (\ref{flateigenvalues}) is
also valid for massless fermions.

\subsection{~Flat Spacetime}
Let us now consider the Casimir energy density in the conformally 
related flat space problem. We shall first look at the effective potential 
per unit area on the brane, ${\cal A}$. For bosons, this 
is given 
\begin{equation}
V^b_0 = {1\over 2 {\cal A}} {\rm Tr}\ {\rm\ln} (-\bar\Box^{(0)}/\mu^2).
\end{equation}
Here $\mu$ is an arbitrary renormalization scale.  Using zeta function
regularization (see e.g. \cite{ramond}), it is straightforward to show that
\begin{equation} V^b_0 (L)= {(-1)^{\eta-1} \over (4\pi)^{\eta} \eta!}
    \left({\pi\over L}\right)^{D-1} \zeta'_R(1-D).
\label{vboson}
\end{equation}
Here $\eta=(D-1)/2$, and $\zeta_R$ is the standard Riemann's zeta function.
The contribution of a massless fermion is given by the same expression 
but with opposite sign:
\begin{equation}
V^{f}_0(L) = - V_0^b(L).
\label{vfermion}
\end{equation}
The expectation value of the energy momentum tensor
is traceless in flat space for conformally invariant 
fields. 
Moreover, because of the symmetries of
our background, it must have the form \cite{bida}
$$
\langle T^z_{\ z}\rangle_{flat}= (D-1) \rho_0(z),\quad 
\langle T^i_{\ j}\rangle_{flat}= -{\rho_0(z)}\ \delta^i_{\ j}.
$$
By the conservation of energy-momentum, $\rho_0$ must
be a constant, given by
$$
\rho_0^{b,f} = {V_0^{b,f} \over 2 L} = \mp {A \over 2 L^D},
$$
where the minus and plus signs refer to bosons and fermions respectively and we
have introduced
$$
A\equiv{(-1)^{\eta} \over (4\pi)^{\eta} \eta!}
\pi^{D-1} \zeta'_R(1-D) > 0.
$$
This result \cite{adpq,dpq}, which is a simple generalization to
codimension-1 branes embedded in higher dimensional spacetimes of the usual
Casimir energy calculation, and it reproduces the same kind of behaviour: the
effective potential depends on the interbrane distance monotonously. So,
depending on $D$ and the field's spin, it induces an atractive or repulsive
force, describing correspondingly the collapse or the indefinite separation of
the branes, just as happened in the Appelquist and Chodos calculation \cite{ac}.
In this case, then, the stabilization of the interbrane distance cannot be due
to quantum fluctuations of fields propagationg into the bulk.

\subsection{~AdS Spacetime}
Now, let us consider the curved space case.  Since the bulk dimension is odd,
there is no conformal anomaly \cite{bida} and the energy momentum tensor is
traceless in the curved case too.\footnote{One can see that for the
  conformally coupled case and for the spacetime we are considering the anomaly
  vanishes even on the branes. Since the conformal anomally is given by the
  Seeley-de Witt coefficient $a_{5/2}$, and this is a conformal invariant
  quantity, in the conformally coupled case $a_{5/2}$ is the same as the flat
  spacetime related problem, which is zero in this case because the branes are
  flat. We can also use the expressions found for $a_{5/2}$ presented in Ref.
  \cite{klaus,vass} to show how in this case it vanishes.}  This tensor is related
to the flat space one by (see e.g.  \cite{bida})
$$
<T^{\mu}_{\ \nu}>_g = a^{-D} <T^{\mu}_{\ \nu}>_{flat}.
$$
Hence, the energy density is given by
\begin{equation}
\rho = a^{-D} \rho_0.
\label{dilute}
\end{equation}
The effective potential per unit physical volume on the positive tension
brane is thus given by
\begin{equation}
V_{\hbox{\footnotesize\it \hspace{-6pt} eff\,}}(\lambda) =
2\ a_+^{1-D} \int a^D(z) \rho\, dz = 
\mp \ell^{1-D}{A \lambda^{D-1} \over (1-\lambda)^{D-1}}.
\label{ve1}
\end{equation}
Note that the background solution $a(z)=\ell/z$ has only been used in 
the very last step. 

\begin{figure}[t]
\centering
\hspace*{-4mm}
%\leavevmode
\epsfysize=10 cm 
\epsfbox{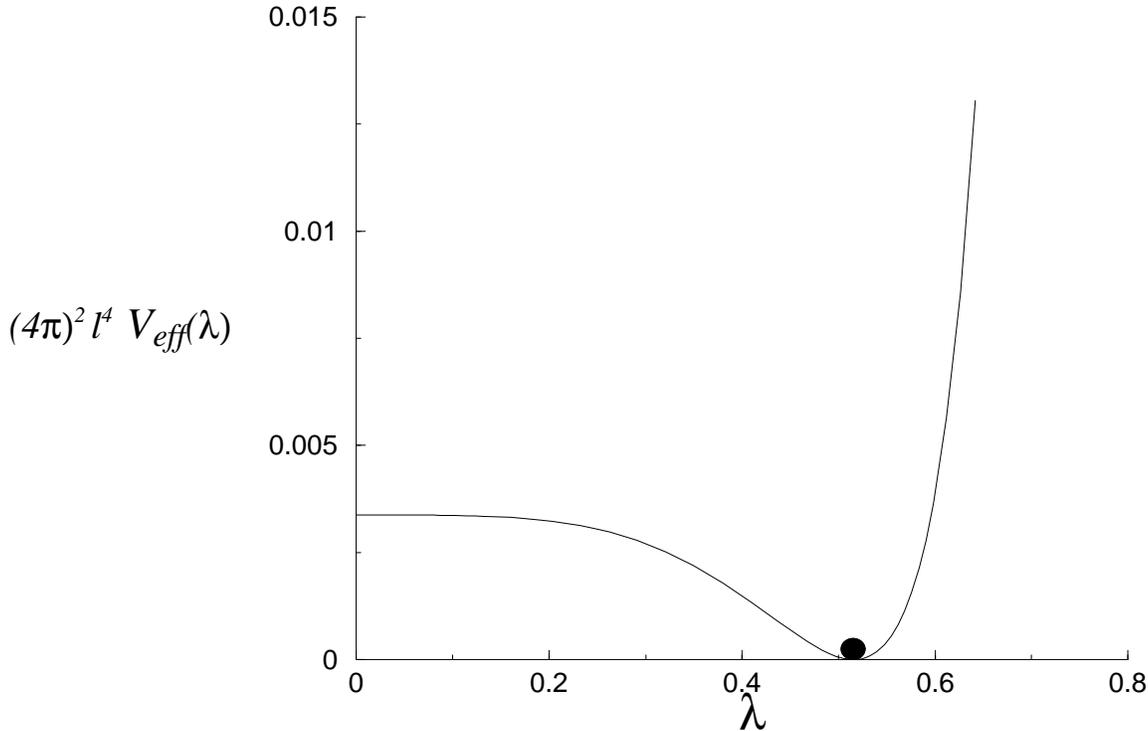}\\[3mm]
%\vspace*{6cm}
\caption{Contribution to the radion 
effective  potential from a massless bulk fermion. This is plotted 
as a function of the dimensionless radion  $\lambda=e^{-d/\ell}$,
where $d$ is the physical interbrane distance. The 
renormalization conditions (\ref{renc}) have been imposed in order to
determine the coefficients $\alpha$ and $\beta$ which appear in 
(\ref{confveff}).}
\label{fig1}
\end{figure}

The previous expression for the effective potential takes into account the
casimir energy of the bulk, but it is not complete because in general the
effective potential receives additional contributions from both branes.
%As we shall discuss in more 
%detail in Section IV, 
We can always add to $V_{\hbox{\footnotesize\it \hspace{-6pt} eff\,}}$ terms
which correspond to finite renormalization of the tension on both branes. These
are proportional to $\lambda^0$ and $\lambda^{D-1}$.  The coefficients in front
of these two powers of $\lambda$ cannot be determined from our calculation and
can only be fixed by imposing suitable renormalization conditions which relate
them to observables.  Adding those terms and particularizing to the case of
$D=5$, we have
\begin{equation}
V_{\hbox{\footnotesize\it \hspace{-6pt} eff\,}}(\lambda) = \mp \ell^{-4}\left[{A\lambda^4 \over (1-\lambda)^4} +
\alpha+\beta\lambda^4\right],
\label{confveff}
\end{equation}
where $A\approx 2.46 \cdot 10^{-3}$.  The values $\alpha$ and $\beta$ can be
obtained from the observed value of the ``hierarchy'', $\lambda_{obs}$, and the
observed value of the effective four-dimensional cosmological constant, which we
take to be zero.  Thus, we take as our renormalization conditions
\begin{equation}
V_{\hbox{\footnotesize\it \hspace{-6pt} eff\,}}(\lambda_{obs})
={dV_{\hbox{\footnotesize\it \hspace{-6pt} eff\,}}\over d\lambda}(\lambda_{obs})=0. 
\label{renc}
\end{equation}
If there are other bulk fields, such as the graviton, which give additional
classical or quantum mechanical contributions to the radion potential, then
those should be included in $V_{\hbox{\footnotesize\it \hspace{-6pt} eff\,}}$.
From the renormalization conditions (\ref{renc}) the unknown coefficients
$\alpha$ and $\beta$ can be found, and then the mass of the radion is
calculable. In Fig.~\ref{fig1} we plot (\ref{confveff}) for a fermionic field
and a chosen value of $\lambda_{obs}$.

From (\ref{renc}), we have
\begin{equation}
\beta = - A (1-\lambda_{obs})^{-5},\quad \alpha= -\beta 
\lambda_{obs}^5.
\label{consts}
\end{equation}
These values correspond 
to changes $\delta \sigma_{\pm}$ on the positive and negative
brane tensions which are related by the equation
\begin{equation}
\delta\sigma_+ =  -\lambda^5_{obs}\ \delta\sigma_-.
\label{reltensions}
\end{equation}
As we shall see below, Eq. (\ref{reltensions}) is just 
what is needed in order to have a static solution according
to the five dimensional equations of motion, once the casimir 
energy is included.

We can now calculate the 
mass of the radion field $m_\phi^{(-)}$ from the point of view of the negative 
tension brane. For $\lambda_{obs}\ll 1$ we have:
\begin{equation}
m^{2\ (-)}_\phi = 
\lambda_{obs}^{-2}\ m^{2\ (+)}_\phi = \lambda_{obs}^{-2}\ 
   {d^2 V_{\hbox{\footnotesize\it \hspace{-6pt} eff\,}}\over
  d\phi^2}\approx \mp
\lambda_{obs} \left({5\pi^3 \zeta'_R(-4)\over 6 M^3 l^5}\right).
\label{massconf}
\end{equation}
The contribution to the radion mass squared is negative for bosons and 
positive for fermions. Thus, depending on the matter content of the 
bulk, it is clear that the radion may be stabilized due to this effect.

Note, however, that if the ``observed'' interbrane separation is large,
then the induced mass is small. So if we try to solve the
hierarchy problem geometrically with a large internal volume, 
then $\lambda_{obs}$ is of order $TeV/m_{pl}$ and the mass 
(\ref{massconf}) is much smaller than the $TeV$ scale. Such a light 
radion would seem to be in conflict with observations.  
In this case we must accept the existence of 
another stabilization mechanism (perhaps classical or 
nonperturbative) contributing a large mass to the radion. 
Of course, another possibility is to have $\lambda_{obs}$ of order one, 
with $M$ and $\ell$ of order $m_{pl}$, in which case the radion mass
(\ref{massconf}) would be very large, but then we must look for a 
different solution to the hierarchy problem. 

\subsection{~Casimir Energy Backreatcion}
Due to conformal invariance, it is straightforward to take into account the 
backreaction of the Casimir energy on the geometry. First of all, we note 
that the metric (\ref{rsmetric}) is analogous to a
Friedmann-Robertson-Walker metric, where the nontrivial direction is space-like
instead of timelike. The dependence of $a$ on 
the transverse direction can be found from the Friedmann equation
\begin{equation}
\left({a'\over a}\right)^2 = {16\pi G_5 \over 3} \rho - {\Lambda \over 6}.
\label{friedmann}
\end{equation}
Here a prime indicates derivative with
respect to the proper coordinate $y$ [see Eq. (\ref{rsmetric})], 
and $\Lambda<0$ is
the background cosmological constant.
Combined with (\ref{dilute}), which relates the energy density $\rho$
to the scale factor $a$, Eq. (\ref{friedmann}) becomes a first order 
ordinary differential equation for $a$. We should also take into account the
matching conditions at the boundaries
\begin{equation}
\left({a'\over a}\right)_{\pm}={\mp 8\pi G_5 \over 6} \sigma_{\pm}. 
\label{matching}
\end{equation}
A static solution of Eqs. (\ref{friedmann}) and (\ref{matching}) can
be found by a suitable adjustment of the brane tensions. Indeed, since the 
branes are flat, the value of the scale factor on the positive tension
brane is conventional and we may take $a_+=1$. Now, the tension $\sigma_+$
can be chosen quite arbitrarily. Once this is done, 
Eq. (\ref{matching}) determines the derivative $a'_+$, and
Eq. (\ref{friedmann}) determines the value of $\rho_0$.
In turn, $\rho_0$ determines the 
co-moving interbrane distance $L$, and hence the location of the second brane.
Finally, integrating (\ref{friedmann}) up to the second brane, 
the tension $\sigma_-$  must be adjusted so that the matching 
condition (\ref{matching}) is satisfied. 
Thus, as with other stabilization scenarios, a single fine-tuning is 
needed in order to obtain a vanishing four-dimensional cosmological constant.

This is in fact the dynamics underlying our 
choice of renormalization conditions (\ref{renc}) which we used in order
to determine $\alpha$ and $\beta$. 
Indeed, let us write
$\sigma_+=\sigma_0 + \delta\sigma_+$ and $\sigma_- =-\sigma_0 +\delta\sigma_-$,
where $\sigma_0=(3 / 4\pi \ell G_5)$ is the absolute value of the 
tension of the branes in the zeroth order background solution. Elliminating
$a'/a$ from (\ref{matching}) and (\ref{friedmann}), we easily recover 
the relation (\ref{reltensions}), which had previously been obtained by 
extremizing the effective potential and imposing zero effective 
four-dimensional cosmological constant (here, $\delta\sigma_{\pm}$
is treated as a small parameter, so that extremization of the effective
action coincides with extremization of the effective potential on the 
background solution.) In that picture, the necessity of a single fine 
tuning is seen as follows. The tension on one of the walls can be chosen 
quite arbitrarily. For 
instance, we may freely pick a value for $\beta$, which renormalizes 
the tension of the brane located at $z_-$. Once this is given, 
the value of the interbrane distance $\lambda_{obs}$ is fixed by the first of
Eqs. (\ref{consts}). Then, the value of $\alpha$, which renormalizes 
the tension of the brane at $z_+$, must be fine-tuned to satisfy the second of
Eqs. (5.8).

Eqs. (\ref{friedmann}) and (\ref{matching}) can of course be solved 
nonperturbatively. We may consider, for instance, the situation where 
there is no background cosmological constant ($\Lambda=0$). In this case 
we easily obtain
\begin{equation}
a^3(z)={6\pi G A \over (z_- -z_+)^5}(C-z)^2
={{3\over 4}\pi^3 \zeta'_R(-4) G_5 }{(z_0-z)^2\over(z_- -z_+)^5},
\end{equation}
where the brane tensions are given by 
$$
2\pi G \sigma_{\pm}=\pm (C-z_{\pm})^{-1}
$$
and $C$ is a constant.
This is a self--consistent solution where the warp in the extra dimension
is entirely due to the Casimir energy.

% Aquesta {\'e}s una soluci{\'o} autoconsistent de les equacions d'Einstein
% semicl{\`a}ssiques on el factor d'escala no trivial {\'e}s degut purament a
% l'energia de Casimir de les pr{\`o}pies branes.
Of course, the conformal interbrane distance $(z_--z_+)$ is different from the
physical $d$, although they are related. For instance, imposing $a(z_+)=1$, which we
can rewrite  as
$$
6\pi A G_5 = \left({z_--z_+\over z_0-z_+}\right)^2 (z_--z_+)^3 
$$
and we get the relation
$$
d=(z_--z_+)\left[{3\over5}\sqrt{(z_--z_+)^3 \over 6\pi G_5 A}\left(
1-\biggl(1-\sqrt{6\pi G_5 A\over(z_--z_+)^3} \;\biggr)^{5/3}\right)\right].
$$
%o b{\'e}
%\begin{equation}
%\left(1-\sqrt{6\pi G_5 A\over
%      {(z_--z_+)^5}}\;(z_--z_+)\right)=\left(1-{5\over3}\sqrt{6\pi G_5 A\over
%      {(z_--z_+)^5}}\;d \right)^{3/5},
%\end{equation}
%Here we can see that when the effect of the Casimir energy is small (and so is the
%curvature consequently), 
Here we can see that the case of negligible Casimir energy, 
${6\pi G_5 A/(z_--z_+)^3} \ll 1$, 
indeed corresponds to the flat case, in which the conformal and the physical
distances coincide.

We can also integrate Eq. (\ref{friedmann}) in the general case \cite{tesina},
and get
\begin{equation}
a(y)=\left({16\pi A M^3 \over{-\Lambda 
        (z_--z_+)^5}}\right)^{1/5}\sinh^{2/5}\left({5\over2}\sqrt{-\Lambda/6}\;(y_0-y)\right),
\end{equation}
with brane tensions given by
$$
\sigma_{\pm}=\pm
{3\over{4\pi}}{\sqrt{-\Lambda/6}\over
  G_5}\coth\left({5\over2}\sqrt{-\Lambda/6}\;(y_0-y_\pm)\right).
$$
Here we are assuming
$\Lambda<0$, and $y_0$ 
is an integration constant.
Moreover  we can explicitely check how this  reduces to RS solution in the
limit of small Casimir energy compared to the cosmological constant, {\em i.e.},
when
$$
{16\pi G_5 \over 3} \rho_0 \ll {\Lambda \over 6}.
$$
Again fixing $a(z_+)=1$ we find
$$
y_0={2\over5}\sqrt{-6\over\Lambda}\; {\rm arcsinh}\left(\left({32\pi
  \rho_0\over-\Lambda M^3}\right)^{-1/2}\right) >> 1
$$
since $(32\pi \rho_0/(-\Lambda M^3)) << 1 $, so that we can write the warp
factor as a power series in the parameter
$(32\pi \rho_0/(-\Lambda M^3))^{1/5} \ll 1$:
$$
a(y)\approx e^{-\sqrt{-\Lambda/6}\;y}\left( 1 - {1\over5}\left({128\pi
      \rho_0 \over -\Lambda M^3}\right)^{2/5} e^{2\sqrt{-\Lambda/6}\;y} +\dots
\right).
$$

\section{Conclusions and discussion}

We have shown that in brane-world scenarios with a warped extra dimension, 
it is in principle possible to stabilize the radion $\phi$ through the Casimir
force induced by bulk fields. Specifically, conformally
invariant fields induce an effective potential of the form (\ref{confveff}) as
measured from the positive tension brane. From the point of view of the 
negative tension brane, this corresponds to an energy density per unit 
physical volume of the order
$$
V_{\hbox{\footnotesize\it \hspace{-6pt} eff\,}}^{-}\sim m_{pl}^4
\left[{A\lambda^4 \over (1-\lambda)^4}+\alpha+\beta\lambda^4\right],
$$
where $A$ is a calculable number (of order $10^{-3}$ per degree of freedom),
and $\lambda \sim \phi/(M^3 \ell)^{1/2}$ is the dimensionless radion. Here
$M$ is the higher-dimensional Planck mass, and $\ell$ is the AdS radius, 
which are both assumed to be of the same order, whereas $m_{pl}$ is the 
lower-dimensional Planck mass. In the absence of any fine-tuning,
the potential will have an extremum at $\lambda \sim 1$,
where the radion may be stabilized (at a mass of order $m_{pl}$).
However, this stabilization scenario without fine-tuning would not explain
the hierarchy between $m_{pl}$ and the $TeV$.

A hierarchy can be generated by adjusting $\beta$
according to (\ref{consts}), with $\lambda_{obs}\sim (TeV/m_{pl}) \sim 
10^{-16}$ (of course one must also adjust $\alpha$ in order to have vanishing
four-dimensional cosmological constant). But with these adjustement,
the mass of the radion would be very small, of order
\begin{equation} 
m^{2\ (-)}_{\phi} \sim\lambda_{obs}\ M^{-3} \ell^{-5} 
\sim \lambda_{obs} (TeV)^2.
\label{smallmass}
\end{equation}
Therefore, in order to make the model compatible with observations, 
an alternative mechanism must be invoked in order to  stabilize the radion,
giving it a mass of order $TeV$. 

Goldberger and Wise \cite{gw1,gw2}, for instance, introduced a field $v$ with
suitable classical potential terms in the bulk and on the branes. In this model,
the potential terms on the branes are chosen so that the v.e.v. of the field in
the positive tension brane $v_+$ is different from the v.e.v. on the negative
tension brane $v_-$. Thus, there is a competition between the potential energy
of the scalar field in the bulk and the gradient which is necessary to go from
$v_+$ to $v_-$. The radion sits at the value where the sum of gradient and
potential energies is minimized. This mechanism is perhaps somewhat {\em ad
  hoc}, but it has the virtue that a large hierarchy and an acceptable radion
mass can be achieved without much fine tuning. It is reassuring that in this
case the Casimir contributions, given by (\ref{smallmass}), would be very small
and would not spoil the model.

The graviton contribution to the radion effective potential can be computed as
well. Each polarization of the gravitons contribute as minimally coupled massles
bulk scalar field \cite{tamaproof}, so since gravitons are not conformally
invariant, the calculation is considerably more involved, and a suitable method
has been developed for this purpose \cite{gpt}.  
% This consists basically of two
% steps. First, one computes the determinant of the operator of the flat space
% (conformally) related problem. One needs the techniques developed in
% \cite{lr1,lr2,elr} in order to sum the tower of Kaluza-Klein masses, defined
% implicitely in this case.  Then one can evluate the difference of this and the
% deteminant of the original operator \cite{bg,mks,dw} by
% integrating along any path that connects both metrics the appropriate
% Seeley-DeWitt coefficient, $a_{5/2}$, recently computed in \cite{klaus,bgkv}.
The result is that gravitons contribute
a negative term to the radion mass squared, but this term is even smaller than
(\ref{smallmass}), by an extra power of $\lambda_{obs}$.
More over this method works also in AdS
space for scalar fields of any kind (massive, nonminimally coupled \dots).

In an interesting recent paper \cite{FH}, Fabinger and Ho\v rrava have considered
the Casimir force in a brane-world scenario similar to the one discussed 
here, where the internal space is topologically $S^1/Z_2$. 
In their case, however, the gravitational field of 
the branes is ignored and the extra dimension is not warped. As a result, 
their effective potential is monotonic and stabilization does not occur 
(at least in the regime where the one loop calculation is reliable, just 
like in the original Kaluza-Klein compactification on a circle \cite{ac}). 
The question of gravitational backreaction of the
Casimir energy onto the background geometry is also discussed 
in \cite{FH}. Again, since the gravitational field of the branes is not 
considered, they do not find static solutions. This is in contrast with our
case, where static solutions can be found by suitable adjustment of the brane 
tensions. 

Finally, it should be pointed out that the treatment of backreaction
(here and in \cite{FH}) applies to conformally invariant fields but not to 
gravitons. Gravitons are similar to minimally coupled scalar 
fields, for which it is well known that the 
Casimir energy density diverges near the boundaries 
\cite{bida}. Therefore, a physical cut-off related 
to the brane effective width seems to be needed so that the energy density 
remains finite everywhere. Presumably, our conclusions will be 
unchanged provided that this cut-off length is small compared with the 
interbrane separation, but further investigation of this issue 
would be interesting.

It seems also interesting to clarify whether the same stabilization mechanism
works in other kind of warped compactified brane world models, such as some
coming form M-theory \cite{ovrut}. In this case the bulk instead of a slice of AdS
(which is maximally symmetric),  consists of a power-law warp factor, and
consequently a less symmetric space. This complicates the calculation since, for
instance, there are two 4-d massless moduli fields (apart from the 4-d
gravitons) to stabilize.

After the work reported here \cite{gpt,tesina} was complete, 
Ref. \cite{muko} appeared with some overlapping results, and also \cite{bmno,noz}
with related topics.

% In contrast with this reference,
% we allow the conformally coupled field in the bulk to be fermionic only, in
% order to obtain stable static solutions.

\section{Acknowledgements}

% We are grateful to Klaus Kirsten for useful discussions.
% J.G. and O.P. acknowledge support from CICYT, under grant
% AEN99-0766. 
% O.P. 
I devote special thanks to Jaume Garriga and Takahiro Tanaka for the entire
realization of this work. I would like to thank Enric Verdaguer  and Edgar Gunzig 
for his kind hospitallity in the Peyresq-5 Meeting.  I acknowledge support from
CICYT, under grant AEN99-0766, and from grant 1998FI 00198.  
%The stay of
%Takahiro Tanaka at IFAE was supported from Monbusho System to Send Researchers
%Overseas.

\end{document}